# Systematic investigation of electrical contact barriers between different electrode metals and layered GeSe

Ranran Li,[1, 2, 3] Wei Xia,[1] Yanfeng Guo,[1] and Jiamin Xue[1a)]

[1] *School of Physical Science and Technology, ShanghaiTech University, Shanghai 201210, China*

[2] *University of Chinese Academy of Sciences, Beijing 100049, China*

[3] *Shanghai Institute of Ceramics, Chinese Academy of Sciences, Shanghai 200050, China*

For electronic and photoelectronic devices based on GeSe, an emergent two dimensional monochalcogenide with many exciting properties predicted, good electrical contacts are of great importance for achieving high device performances and exploring the intrinsic physics of GeSe. In this article, we use temperature-dependent transport measurements and thermionic emission theory to systematic investigate the contact-barrier heights between GeSe and six common electrode metals, Al, Ag, Ti, Au, Pt and Pd. These metals cover a wide range of work functions (from ~ 3.6 eV to ~ 5.7 eV). Our study indicates that Au forms the best contact to the valence band of GeSe, even though Au does not possess the highest work function among the metals studied. This behavior clearly deviates from the expectation of Schottky-Mott theory and indicates the importance of the details at the interfaces between metals and GeSe.

GeSe, a member of group IV monochalcogenides, is a layered semiconductor attracting increasing attentions.[1-3,4] Its direct and indirect bandgaps covers the entire solar spectrum, making it a potential candidate for photodetecting and photovoltaic applications with good energy conversion efficiency.[2,4,5] Its puckered lattice structure, similar to that of black phosphorus (BP), gives rise to in-plane anisotropic response to external stimulations such as polarized light illumination.[6] In addition to the many experimental efforts, theoretical studies have unveiled even more exciting properties. First-principle calculations showed that the thermoelectric figure of merit (ZT) of hole doped GeSe can reach as high as 2.6, even higher than that of SnSe, the experimentally confirmed material with the record high ZT.[7] Due to broken inversion symmetry, monolayer GeSe was predicted to have a huge piezoelectric coefficient which is one to two magnitude larger than traditional piezoelectric materials like AlN.[8] The same broken symmetry also produces a spontaneous polarization in monolayer GeSe which is expected to generate large photocurrent (shift current) even without a p-n junction.[9]

To fully realize its great potential, devices based on two-dimensional (2D) GeSe mainly in the form of the field-effect transistors (FETs) are needed. For this type of device, good electrical contacts between the electrodes and GeSe with low Schottky barrier height (BH) are crucial for studying the intrinsic properties of GeSe and minimizing the influence of metal-semiconductor contacts. In this work, we use temperature-dependent transport measurements to experimentally determine the contact BH between exfoliated few-layer GeSe flakes and six common electrode materials, namely Ti, Al, Ag, Pd, Au and Pt. These metals cover a wide range of work functions (WF) from ~ 3.6 eV (Ti) to ~ 5.7 eV (Pt). According to the ideal Schottky-Mott theory, the BH depends solely on WF difference between a metal and the semiconductor.[10] So we expect Pt to form the best contact with the valence band of GeSe. However, we find that Au electrodes have the lowest BH to GeSe. This study demonstrates the importance of interface details at determining the contact barriers and suggests the choice of electrode material for GeSe devices, which is of great importance for future experimental study of GeSe.

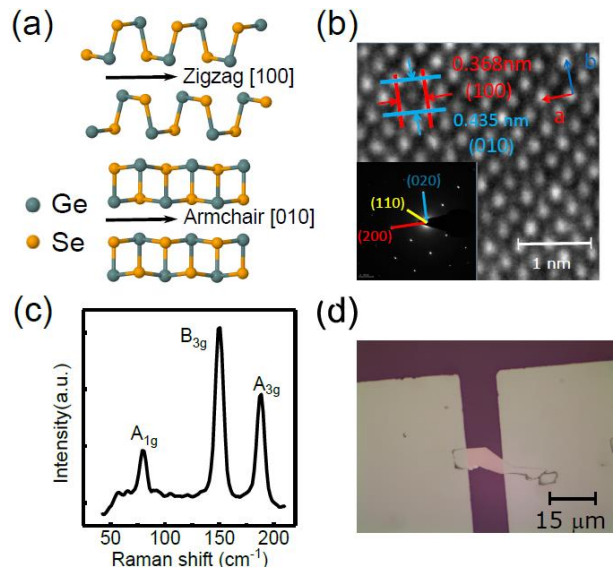

FIG. 1. (a) Atomic arrangement of bulk GeSe along the armchair and zigzag directions. (b) TEM image of exfoliated GeSe. The lattice constants along [100] and [010] directions are 0.383 nm and 0.438 nm, respectively. The inset is the SAED pattern corresponding to the TEM image. (c) Raman spectrum of a GeSe thin flake. (d) The optical microscopy image of a GeSe FET device with channel thickness of 40 nm.

Bulk GeSe belongs to the orthorhombic space group *Pnma* (No. 62). Its crystal has a similar structure to that of BP as shown in Figure 1(a). The two principle axes along [100] and [010] are named as the zigzag and armchair directions, respectively. Unlike BP, which is an elemental crystal, GeSe consists of two elements with different electronegativity. This breaks the inversion symmetry in odd numbered layers. The GeSe crystals used in this study are grown by the chemical-vapor-transport method (see the Supplementary Material, SM, for growth details) with elementary germanium and selenium as the starting materials. Typical crystal sizes are a few millimeters. Scotch-tape method is used to exfoliate GeSe onto degenerately doped Si substrate covered with 300 nm $SiO_2$. To confirm the crystallinity of our sample, characterizations with transmission-electron microscopy (TEM) and Raman spectroscopy are performed. The atomically resolved TEM image in Figure 1(b) of GeSe flakes transferred to the TEM grid clearly shows the orthorhombic lattice in the (001) plane. The measured lattice constants are in close agreement with previously reported values.[11] Defects are seldom observed at various locations of the sample, and selected-area electron diffraction (SAED, inset of Figure 1(b)) over a large sample area consists of a single set of sharp diffraction points. Raman spectrum (Figure 1(c)) on samples exfoliated onto $SiO_2$ also reveals the characteristic phonon modes of GeSe with strong Raman peaks.[4] All these evidences demonstrate the high crystallinity of our GeSe and the stability of thin-flake GeSe under ambient condition.

To study transport properties of GeSe and evaluate the BH between different electrode metals and GeSe, FETs based on thin flakes of GeSe are fabricated. Conventionally, this type of devices is fabricated with electron-beam lithography, metal deposition and lift-off process. However, this method put constraints on the choice of metals. For example, the most commonly used Au needs a wetting layer of Ti at the bottom due to the poor adhesion between Au and $SiO_2$. Otherwise, in the lift-off process Au would easily come off the $SiO_2$ sample. However, a thin layer of Ti would completely change the interface between Au and GeSe, hence no intrinsic BH could be extracted. To avoid this complication, we use a facile shadow mask technique. A TEM grid working as a shadow mask is transferred onto the GeSe flake and electrode metal is deposited through the grid. The optical microscopy image of a typical device is shown in Figure 1(d). This method produces clean two-terminal devices without the contamination of photoresist and has no metal adhesion issue. Only when we need to perform four-probe transport measurement we use the conventional microfabrication techniques to fabricate multi-terminal devices.

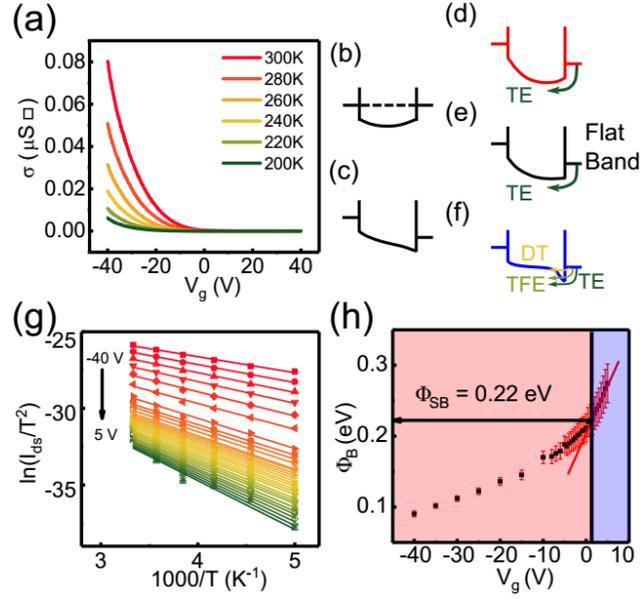

FIG. 2. (a) Transfer characteristics of a back-gated GeSe device with Ti contacts at various temperatures. (b)~(f) Band-alignment diagrams at thermal equilibrium (b), large drain-source bias (c), thermionic emission (d), flat band (e) and tunneling (f) conditions. (g) Arrhenius-type plot at different $V_g$ from -40 V to 5 V to extract the contact BH $\Phi_B$. (h) Calculated BH $\Phi_B$ as a function of gate voltage $V_g$. The Schottky BH between GeSe and Ti is $\Phi_{SB} = 0.22$ eV.

We first examine the contact between Ti and GeSe as shown in Figure 2, as Ti is one of the most commonly used electrode metals for fabricating devices based on 2D materials. We have studied multiple devices and obtained similar results (data from other Ti contacted devices are presented in Figure S1(b) and Table S1 in the SM). Temperature-dependent two-terminal transport measurements in Figure 2a clearly indicate that the GeSe FET is p type, consistent with previous results of bulk GeSe and thin flakes.[1,12,13] Since no intentional dopants are introduced in crystal growth, this doping has its intrinsic origin, most likely from Ge deficiency. Calculations of similar materials, such as SnSe,[14] have indicated that Sn vacancy has a moderate formation energy (less than 1 eV, to be compared with growth temperature of 800 °C). This intrinsic p-type doping is the common feature of group IV monochalcogenides.[15] As temperature goes down, holes are frozen out, which results in dropping conductance together with a negative shift of threshold voltage as expected.

From the temperature-dependent transfer curves, the contact BH of titanium and GeSe can be extracted based on the band diagrams shown in Figures 2b ~ f.[16] Without the applied drain-source bias, the FET consists of two back-to-back Schottky diodes with possible band bending near the GeSe-electrode interfaces (Figure 2b). At bias close to 0 V, charge carriers (holes for this case) need to overcome both barriers at the two interfaces to reach from the source to drain (in this study the drain electrode is the one kept grounded). In order to get information about one barrier, a larger voltage of 5 V is applied between the source and drain electrodes to forward bias the Schottky diode at the drain side (Figure 2c). We note that due to a long channel (several microns) of our device, drain-induced barrier lowering is negligible.[17] In this way, the subthreshold current is mainly controlled by the barrier at the source contact, which in turn is modulated by the gate voltage $V_g$. When $V_g$ is positive (Figure 2d), the band edges in the channel are pulled down in energy, which increases the BH for holes and the device is turned off. As the $V_g$ decreases, a flat-band condition is reached (Figure 2e). The intrinsic BH is measured at this point. Since the barrier is very wide (in the micron range), holes cannot tunnel through it to reach the drain electrode. Only thermionic emission (TE) can contribute to the current (represented by the arrows in Figures 2d and 2e). When the $V_g$ is increased in the negative direction, the band edges in the channel are pushed up in energy. The barrier width at the source side is thinned down (Figure 2f) and tunneling, including direct tunneling (DT) and thermally-assisted tunneling (i.e. thermal-field emission, TFE), becomes progressively easier. Then the effective BH is reduced as a function of $V_g$. Based on this analysis, we expect the following behavior of the effective BH as a function of $V_g$ extracted from the temperature-dependent transport data in Figure 2a: the measured barrier will be higher than the intrinsic value at large positive gate voltage, linearly decreases as the $V_g$ is reduced, and finally it will depart from the linear trend with a sublinear behavior after passing the flat-band point (this is because that the metal Fermi levels are pinned related to the band edges at the contact interfaces as shown in Figure 2f, resulting in a constant BH for thermionic emission alone).

Current due to thermionic emission can be expressed as: [17]

$$I_{ds} = AT^2\exp(-\frac{q\phi_B}{k_BT})[1-\exp(-\frac{qV_{ds}}{k_BT})],$$

where $A$ is the Richardson's constant, $T$ is the temperature, $q$ is the elementary charge, $\phi_B$ is the contact BH, $k_B$ is the Boltzmann constant and $V_{ds}$ is the source-drain voltage. Since $V_{ds}$ is much higher than $k_BT$, the equation can be simplified to $I_{ds} = AT^2\exp(-\frac{q\phi_B}{k_BT})$. We then use this formula to fit the temperature dependence of $I_{ds}$ at a certain $V_g$ to obtain the effective BH (Figure 2g). The fitted values are plotted as a function of $V_g$ in Figure 2h. As expected, when $V_g$ is swept from the large positive value towards the negative value, the extracted effective BH decreases with a linear regime followed by a sublinear regime. The transition between these two regimes is where the flat-band condition is reached, and the intrinsic BH is obtained. For titanium contacted GeSe, the intrinsic BH for the device in Figure 2 is determined to be 0.223 $\pm$ 0.003 eV and the corresponding flat-band $V_g$ is 1.2 $\pm$ 0.2 V (where the uncertainties are obtained from curve fittings of the data in Figure 2h). More Ti contacted devices are measured with similar results as summarized in Table S1 in the SM. The averaged BH is 0.209 $\pm$ 0.019 V. These values mean that although titanium is commonly used as the contact metal to GeSe,[12] it has a high Schottky barrier for hole transport and hence is not a good choice.

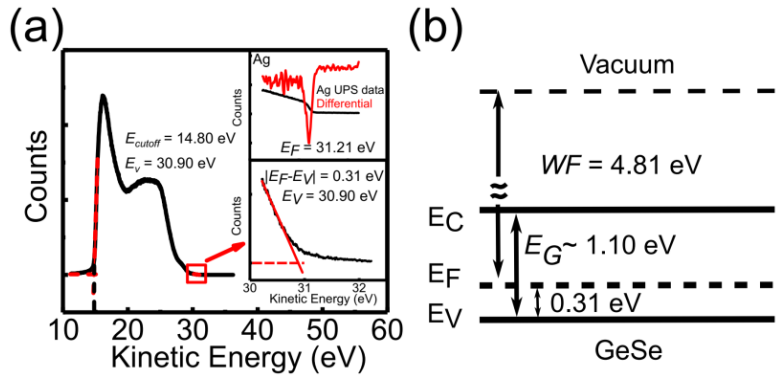

FIG. 3. (a) UPS data to obtain GeSe WF. Upper inset: UPS data of the reference silver to calibrate the Fermi energy. The Fermi energy is at 31.21 eV. Lower inset: Zoom in of the GeSe valence band data. (b) The diagram of GeSe energy levels inferred from the UPS measurement. The WF of GeSe is 4.81 eV and ionization energy is 5.12 eV.

To find a better contact to GeSe, we could use Schottky-Mott theory as a guidance[10]. In this idealized situation, BH between a metal and the valence band of GeSe is the difference between the work function of the metal and the ionization energy of GeSe. We use ultraviolet photoemission spectroscopy (UPS) to measure the energy levels in GeSe as shown in Figure 3. The surface layer of a large GeSe crystal is peeled off right before being inserted into the ultra-high vacuum chamber of UPS, to avoid ambient contamination. Argon-sputtered silver, which is electrically connected with the GeSe crystal, is used to calibrate the Fermi level $E_F$ to be 31.21 eV as shown in the inset of Figure 3a. The WF of GeSe ($\phi$) can be determined by the formula $\phi = h\nu - (E_F - E_{cutoff})$, where $h\nu = 21.22$ eV is the energy of incident photons and $E_{cutoff} = 14.80$ eV is the cutoff energy (a bias of 10 V is applied to accelerate the secondary electron for a better measurement). The WF of GeSe is calculated to be 4.81 eV. By analyzing the onset of photo electrons near the Fermi level, the valence band energy is found to be 0.31 eV below the Fermi energy (Figure 3b). Previous study has found the WF of Ti to be 4.33 eV.[18] Then the BH between Ti and GeSe predicted by Schottky-Mott theory would be ~ 0.79 eV, much higher than the value of 0.209 $\pm$ 0.019 V measured here. Since the WF of a metal film depends on the details of deposition, we have also measured the WF of Ti ourselves with UPS and found it to be 3.58 $\pm$ 0.05 eV (see Table S1 in the SM). Then the theoretical BH is even higher than the measured value. Schottky-Mott theory also indicates that metals with higher WF will form lower BHs with the valence band of GeSe. Au has a higher WF (5.12 ~ 5.93 eV)[18] matching well with the energy levels of GeSe and is another commonly used metal for electrodes. We then fabricate Au contacted devices. Room temperature transfer curves in Figure 4a indicate that the gold contacted device has a two-terminal conductivity one-order-of-magnitude higher than that of the titanium device (see Table S1 in SM for more data about Au contacted devices). Temperature-dependent transport measurements are used to extract the BH versus $V_g$ as shown in Figure 4b. Flat-band condition is reached at $V_g = -18.7 \pm 1.7$ V and BH = $0.100 \pm 0.003$ eV. The averaged BH of five different devices is $0.128 \pm 0.024$ eV, much lower than that of Ti and GeSe, albeit higher than what we expect from Schottky-Mott theory. To get a more complete picture of the contact between different metals and GeSe, we perform similar study on other electrode metals including Al, Ag, Pt and Pd with a wide range of WFs. To rule out the influence of experimental details on the value of WFs, we have measured all the metals films with UPS and the WF values are summarized in Table S1 of the SM. The extracted BHs are ploted in Figure 4c. The red filled circles represent the averaged BHs for each type of electrode metal and the open circles represent the original data for each device. While the WF from Ti to Pt is increasing (see Table S1 in SM for details),[18] the BH does not show monotonic decreasing, as expected by the Schottky-Mott model. Compared with other metals, Au is the best choice for contacting GeSe. Previous study[10] has shown that besides the WF alignment, the interface details also play an important role in determining the Fermi level pinning and hence the BH between a metal and a 2D material. Although Pt has a higher WF compared with that of Au, the evaporation temperature of Pt is significantly higher than that of Au. This harsh deposition condition will introduce more surface defects as shown before[10] and more sever Fermi level pinning is expected. Besides that, the chemistry between the electrodes and semiconductor can also be important. We find that Ag contacted devices show severe alloying between the metal and GeSe during transport measurements, as shown in Figure S4. Hence no reliable devices could be made with Ag electrodes. All these findings indicate the importance of the interface between a metal and the semiconductor. We also note that previous studies (e.g. Reference 18) reported Pd to have a work function of 5.22 to 5.60 eV, but our own measurement has shown a much smaller value (4.69 ± 0.07 eV). This discrepancy is worth further investigation.

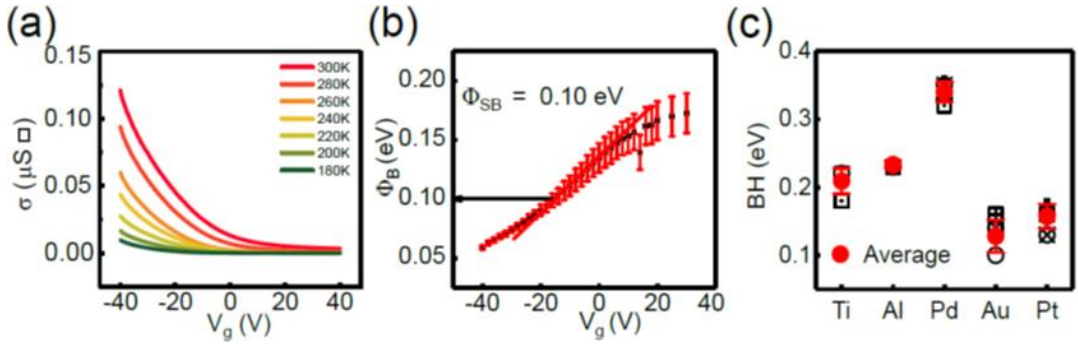

FIG. 4. (a) Transfer curves of GeSe FET contacted with Au measured at different temperatures. (b) Extracted BH $\Phi_B$ of the contact between GeSe and Au versus gate voltage $V_g$. The Schottky BH for Au is $\Phi_{SB} = 0.10$ eV. (c) A systematic investigation of BH of five different metals with increasing WFs from Ti to Pt.

In conclusion, we have measured the Schottky-barrier heights between thin flakes of GeSe and a group of metals – Al, Ag, Ti, Au, Pt and Pd. We have shown that Au is the best choice among the metal studied for contacting GeSe. Through a systematic investigation, we demonstrated that WF alignment is only one of the factors affecting the BH between a metal and a 2D material. Due to the many intriguing electronic properties expected for GeSe,[1,2,3,4,19] our study provides a good guidance for choosing the best electrode metal for GeSe devices.

Supplementary Material

GeSe crystal growth details; Barrier heights of GeSe contacted with different metal electrods; Two-terminal $I_{ds}$-$V_{ds}$ data for GeSe FET devices with different metal contacts; Data of four-terminal and two-terminal transfer curves of a Ti-contacted GeSe device; SEM images of devices with Ag electrodes; Summary of BHs with different metals derived from multiple devices.

R.L and J.X. are supported by the National Natural Science Foundation of China (No. 11504234), Shanghai Municipal Education Commission (No. 15ZZ115), Thousand Talents Program and ShanghaiTech University. W.X. and Y.G. are supported by the Natural Science Foundation of Shanghai (No. 17ZR1443300) and the Shanghai Pujiang Program (No. 17PJ1406200).

1   S. M. Yoon, H. J. Song, and H. C. Choi,  Adv. Mater. **22** (19), 2164 (2010).
2   D. J. Xue, J. Tan, J. S. Hu, W. Hu, Y. G. Guo, and L. J. Wan, Adv. Mater. **24** (33), 4528 (2012).
3   B. Mukherjee, Y. Cai, H. R. Tan, Y. P. Feng, E. S. Tok, and C. H. Sow,  ACS Appl. Mater. Interfaces **5** (19), 9594 (2013); Y.h. Hu, Sh.l. Zhang, Sh.f. Sun, M.q. Xie, B. Cai, and H.b. Zeng,  Appl. Phys. Lett. **107** (12), 122107 (2015);  H. Kim, Y. Son, J. Lee, M. Lee, S. Park, J. Cho, and H. C. Choi,  Chem. Mater. **28** (17), 6146 (2016).
4   H. Zhao, Y.L. Mao, X. Mao, X. Shi, C.S. Xu, C.X. Wang, S. Zhang, and D.H. Zhou,  Adv. Func. Mater. **28** (6), 1704855 (2018).
5   D. J. Xue, S. Ch. Liu, Ch. M. Dai, Sh. Y. Chen, Ch. He, L.  Zhao, J. S. Hu, and L. J. Wan,  J. Am. Chem. Soc **139** (2), 958 (2017).
6   X. Wang, Y. Li, L. Huang, X. W. Jiang, L. Jiang, H. Dong, Z. Wei, J. Li, and W. Hu,  J. Am. Chem. Soc **139** (42), 14976 (2017).
7   L. D. Zhao, G. Tan, S. Hao, J. He, Y. Pei, H. Chi, H. Wang, S. Gong, H. Xu, V. P. Dravid, C. Uher, G. J. Snyder, C. Wolverton, and M. G. Kanatzidis,  Science **351** (6269), 141 (2016).
8   R.X. Fei, W. Li, J. Li, and L. Yang,  Appl. Phys. Lett. **107** (17), 173104 (2015).
9   T. Rangel, B. M. Fregoso, B. S. Mendoza, T. Morimoto, J. E. Moore, and J. B. Neaton,  Phys. Rev. Lett. **119** (6), 067402 (2017);   A. M. Cook, M. Fregoso B, F. de Juan, S. Coh, and J. E. Moore,  Nat. Commun. **8**, 14176 (2017).
10  Y. Liu, J. Guo, E. Zhu, L. Liao, S. J. Lee, M. Ding, I. Shakir, V. Gambin, Y. Huang, and X. Duan,  Nature **557** (7707), 696 (2018).
11  A. Onodera, I. Sakamoto, Y. Fujii, N. Mo, and Sh. Sugai,  Phys. Rev. B **56** (13), 7935 (1997).
12  W. C. Yap, Zh. F. Yang, M. Mehboudi, J.A. Yan, S. Barraza-Lopez, and W.J. Zhu,  Nano Res. **11** (1), 420 (2017).
13  D. D. Vaughn, R. J. Patel, M. A. Hickner, and R. E. Schaak,  J. Am. Chem. Soc. **132** (43), 15170 (2010).
14  Z. Tian, M. Zhao, X. Xue, W. Xia, C. Guo, Y. Guo, Y. Feng, and J. Xue,  ACS Appl. Mater. Interfaces **10** (15), 12831 (2018).
15  O. Madelung, *Semiconductors: Data Handbook*. (Spinger-Verlag, Marburg, 2004).
16  S. Das, H. Y. Chen, A. V. Penumatcha, and J. Appenzeller,  Nano. Lett. **13** (1), 100 (2013).
17  S. M. Sze and K. K. Ng, *Physics of Semiconductor Devices*. (Jonh Wiley & Sons. Hoboken, New Jersey, 2006).
18  H. L. Skriver and N. M. Rosengaard,  Phys. Rev. B **46** (11), 7157 (1992).
19  G. Shi and E. Kioupakis,  Nano. Lett. **15** (10), 6926 (2015);   P. Mishra, H. Lohani, A. K. Kundu, R. Patel, G. K. Solanki, Krishnakumar S. R. Menon, and B. R. Sekhar,  Semicond. Sci. Technol. **30** (7), 075001 (2015);   H. Wang and X.F. Qian,  2D Mater. **4** (1), 015042 (2017).